\documentclass[sigconf]{acmart}
\usepackage{booktabs} 

\copyrightyear{2018} 
\acmYear{2018} 
\setcopyright{acmlicensed}
\acmConference[PEARC '18]{Practice and Experience in Advanced Research Computing}{July 22--26, 2018}{Pittsburgh, PA, USA}
\acmBooktitle{PEARC '18: Practice and Experience in Advanced Research Computing, July 22--26, 2018, Pittsburgh, PA, USA}
\acmPrice{15.00}
\acmDOI{10.1145/3219104.3219135}
\acmISBN{978-1-4503-6446-1/18/07}

\begin{document}
\title{SciTokens: Capability-Based Secure Access to Remote Scientific Data}

\author{Alex Withers}
\affiliation{%
  \institution{NCSA}
}
\email{alexw1@illinois.edu}

\author{Brian Bockelman}
\affiliation{%
  \institution{University of Nebraska-Lincoln}
}
\email{bbockelm@cse.unl.edu}

\author{Derek Weitzel}
\affiliation{%
  \institution{University of Nebraska-Lincoln}
}
\email{dweitzel@cse.unl.edu}

\author{Duncan Brown}
\affiliation{%
  \institution{Syracuse University}
}
\email{dabrown@syr.edu}

\author{Jeff Gaynor}
\affiliation{%
  \institution{NCSA}
}
\email{gaynor@illinois.edu}

\author{Jim Basney}
\orcid{0000-0002-0139-0640}
\affiliation{%
  \institution{NCSA}
}
\email{jbasney@illinois.edu}

\author{Todd Tannenbaum}
\affiliation{%
  \institution{University of Wisconsin-Madison}
}
\email{tannenba@cs.wisc.edu}

\author{Zach Miller}
\affiliation{%
  \institution{University of Wisconsin-Madison}
}
\email{zmiller@cs.wisc.edu}

\renewcommand{\shortauthors}{A. Withers et al.}

\begin{abstract}
The management of security credentials (e.g., passwords, secret keys) for computational science workflows is a burden for scientists and information security officers. Problems with credentials (e.g., expiration, privilege mismatch) cause workflows to fail to fetch needed input data or store valuable scientific results, distracting scientists from their research by requiring them to diagnose the problems, re-run their computations, and wait longer for their results. In this paper, we introduce SciTokens, open source software to help scientists manage their security credentials more reliably and securely. We describe the SciTokens system architecture, design, and implementation addressing use cases from the Laser Interferometer Gravitational-Wave Observatory (LIGO) Scientific Collaboration and the Large Synoptic Survey Telescope (LSST) projects. We also present our integration with widely-used software that supports distributed scientific computing, including HTCondor, CVMFS, and XrootD. SciTokens uses IETF-standard OAuth tokens for capability-based secure access to remote scientific data. The access tokens convey the specific authorizations needed by the workflows, rather than general-purpose authentication impersonation credentials, to address the risks of scientific workflows running on distributed infrastructure including NSF resources (e.g., LIGO Data Grid, Open Science Grid, XSEDE) and public clouds (e.g., Amazon Web Services, Google Cloud, Microsoft Azure). By improving the interoperability and security of scientific workflows, SciTokens 1) enables use of distributed computing for scientific domains that require greater data protection and 2) enables use of more widely distributed computing resources by reducing the risk of credential abuse on remote systems.
\end{abstract}

%
%

\ccsdesc[500]{Security and privacy~Authorization}

\keywords{OAuth, capabilities, distributed computing}

\settopmatter{printfolios=true}
\maketitle

\section{Introduction}

Access control on remote data is an essential capability for computational science. Maintaining the privacy of scientific data prior to publication helps avoid premature science claims and facilitates a healthy competition between research groups. Maintaining the integrity of scientific data (i.e., preventing intentional or unintentional data alteration) helps avoid erroneous scientific claims and preserves the provenance of scientific results. Computational science is often geographically distributed, with scientists collaborating across organizations and using computational resources spread across distributed facilities such as the LIGO Data Grid, the Open Science Grid, XSEDE, and public cloud providers (Amazon, Google, Microsoft, etc.). Secure remote access to scientific data enables these collaborations.

Access control for distributed scientific computing across administrative domains is fundamentally different from local computing environments. When computing on the desktop or a local high performance cluster, the local operating system provides filesystem access control using local accounts, groups, and file permissions. In contrast, when computing across distributed facilities, access control is managed by the collaboration (often called a virtual organization or VO), and access to computing resources and scientific data is determined by membership in that collaboration. A remote compute job, which is often part of a larger computational science workflow, will read input data and write output data to/from remote file servers. These network-accessible file servers may be operated on a long-term basis, serving multiple VOs, or they may be instantiated on-the-fly by the VO (often called a glide-in).

The remote compute job needs security credentials to access those file servers. In common practice today, those credentials are identity tokens, carrying the identity of the individual researcher or the VO, enabling the job to act on behalf of that researcher or VO when accessing remote resources like file servers. Using identity tokens in this way creates significant risk of abuse, since they are used by jobs that are running on remote, less trusted systems and if stolen, these tokens provide wide access to the attacker. The continuous stream of news reports about compromised passwords at sites like LinkedIn and Yahoo! teach us how using the same credential across multiple sites enables attacks to spread widely. The same concern applies to the credentials we use for academic and scientific computing.

To address this risk for current distributed scientific workflows and to enable use of additional distributed computational resources for more scientific domains, SciTokens applies the well-known principle of capability-based security to remote data access. Rather than sending unconstrained identity tokens with compute jobs, we send capability-based access tokens. These access tokens grant the specific data access rights needed by the jobs, limiting exposure to abuse. These tokens comply with the IETF OAuth standard \cite{RFC6749}, enabling interoperability with the many public cloud storage and computing services that have adopted this standard. By improving the interoperability and security of scientific workflows, we 1) enable use of distributed computing for scientific domains that require greater data protection and 2) enable use of more widely distributed computing resources by reducing the risk of credential abuse on remote systems.

\begin{figure}
\includegraphics[scale=0.35]{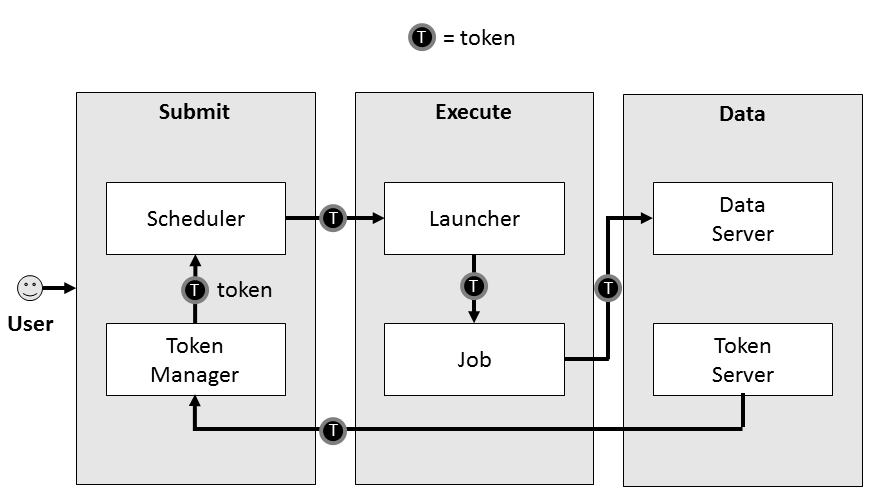}
\caption{The SciTokens Model}
\label{fig:model}
\end{figure}

As illustrated in Figure~\ref{fig:model}, our SciTokens model applies capability-based security to three common domains in the computational science environment: Submit (where the researcher submits and manages scientific workflows), Execute (where the computational jobs run), and Data (where remote read/write access to scientific data is provided). The Submit domain obtains the needed access tokens for the researcher's jobs and forwards the tokens to the jobs when they run, so the jobs can perform the needed remote data access. The Scheduler and Token Manager work together in the Submit domain to ensure that running jobs have the tokens they need (e.g., by refreshing tokens when they expire) and handle any errors (e.g., by putting jobs on hold until needed access tokens are acquired). The Data domain contains Token Servers that issue access tokens for access to Data Servers. Thus, there is a strong policy and trust relationship between Token Servers and Data Servers. In the Execute domain, the job Launcher delivers access tokens to the job's environment, enabling it to access remote data.

\begin{figure}
\includegraphics[scale=0.35]{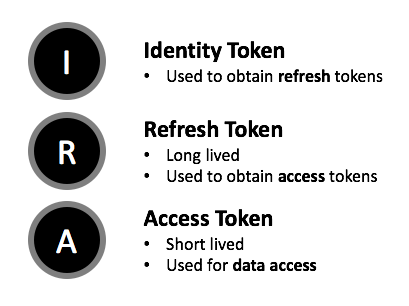}
\caption{Different Token Types}
\label{fig:token_types}
\end{figure}

The SciTokens model adopts token types from OAuth (see Figure~\ref{fig:token_types}). Users authenticate with identity tokens to submit jobs (workflows), but identity tokens do not travel along with the jobs. Instead, at job submission time the Token Manager obtains OAuth refresh tokens with needed data access privileges from Token Servers. The Token Manager securely stores these relatively long-lived refresh tokens locally, then uses them to obtain short-lived access tokens from the Token Server when needed (e.g., when jobs start or when access tokens for running jobs near expiration). The Scheduler then sends the short-lived access tokens to the jobs, which the jobs use to access remote data. 

The remainder of our paper is organized as follows: We first discuss the target use cases for our work. Then we discuss existing capabilities in the software used by LIGO and other science projects that we leverage. Then we detail our technical approach for implementing the SciTokens model. Lastly, we discuss related work, our implementation status, and our conclusions and next steps.

\section{Use Cases}
Our SciTokens work is motivated by use cases from the LIGO, LSST, OSG, and XSEDE projects.

\subsection{LIGO}
The observation of binary black hole mergers \cite{LIGO1,LIGO2,LIGO3} by the Advanced Laser Interferometer Gravitational-wave Observatory (LIGO) \cite{LIGO4} marks a transformative moment for physics and astronomy. Gravitational waves are ripples in the fabric of spacetime produced by the coherent relativistic motion of masses. Gravitational-wave observatories can peer into the cores of exploding stars, study the interiors of neutron stars, and explore the physics of colliding black holes.
The challenge of gravitational-wave detection is to separate the tiny mirror motions caused by gravitational waves from the motion caused by all the other noise sources in the detector. LIGO's searches are most sensitive when we have prior knowledge of the shape of the gravitational waves. In this case, we can use matched filtering to extract the signals from the noisy data. Since we do not know the physical parameters of any given source in advance, we must search for many different sources by matched filtering the data against a large "bank" of gravitational-waveform templates. The noise in the LIGO detectors consists of a stationary, Gaussian component from fundamental sources and non-Gaussian "glitches" of both environmental and instrumental origin. To eliminate glitches from the data, we record a large amount of instrumental health and status information that must be folded into the search. To make confident detections, we demand that a gravitational-wave signal is present in two or more detectors in the network with consistent signal parameters. To measure the significance of a particular candidate signal, we must compare its amplitude to that of the noise-induced background in the network. We measure this background by repeating the search many times with the detector data offset by time intervals larger than the gravitational-wave travel time between the detectors. The statistical significance of candidates is then computed and any significant events are followed-up by additional analysis of auxiliary and environmental detector status channels.

Successfully executing all of the steps described above requires the execution of a search workflow with components that span scientific algorithms, cyberinfrastructure, data management, and distributed computational hardware \cite{Brown06}. A typical gravitational-wave search may generate a workflow of hundreds of thousands of discrete computational tasks, with job dependencies and a large numbers of intermediate data products. The structure of a workflow will vary depending on the input parameters for a specific search. Advanced LIGO uses the "PyCBC" search \cite{PyCBC16} to detect and study gravitational-waves from binary black holes, and to search for binary neutron stars and neutron-star–black-hole mergers. PyCBC-generated workflows are written in an abstract workflow format that can be planned by the Pegasus Workflow Management System \cite{Deelman15} and executed by HTCondor \cite{Thain05} on LIGO Data Grid, Open Science Grid, and XSEDE HTC resources (Comet and Stampede).

Authentication tokens are needed at multiple stages during the creation, planning, and execution of the PyCBC search workflow. A program pycbc\_make\_coinc\_search\_workflow is used to create a workflow to analyze a specific block of data; this is typically two calendar weeks, an interval chosen to allow measurement of the search's noise background to a level where detections can be made. The workflow creation script must make an authenticated query to two metadata servers: the first reports the on the availability and quality of LIGO data, and the second is used to locate the input data files needed. These files may be stored on a local filesystem, on a XrootD server, or available via GridFTP. At execution time, the jobs use an authentication token to access the LIGO data via CVMFS/XrootD or via GridFTP and to fetch any additional workflow data from the submission site via either the GridFTP or scp protocols. The token is then used to push the data from a job back to the submission site for publication (if it is a final data product) or use by other jobs in the workflow (in the case of an intermediate data product).

Analysis of two weeks of data by a PyCBC workflow typically requires approximately 2500 CPU days and takes several wall-clock days to complete. It is not uncommon for intervals of poor data quality (which trigger additional processing in the workflow) to extend the runtime of a workflow to approximately one wall-clock week. At present all authentication during the workflow is performed using a user's X.509 grid proxy credential, which is created with a lifetime that nominally exceeds the execution time of the workflow. This single credential must exist on the submission site and be copied to all of the job execution sites, which can be of order 10,000 individual nodes for a production OSG/XSEDE run. This credential is the same token that can be used to log in to LIGO clusters and authenticate to many LIGO services. If it is compromised, then an attacker can masquerade as the compromised user for the duration of the credential, gaining access to all of the user's files and to all LIGO data.

\subsection{LSST}
LSST's Batch Production service for computing jobs and data processing is still currently being implemented and tested. While the overall design of the system has been architected, there are many details that have to be decided upon \cite{Kowalik16}. LSST has yet to determine authentication and authorization access methods at remote sites. One key stated requirement of LSST's Batch Production service is to provides credentials and endpoint information for any needed LSST services. It is anticipated that access will be in part governed and managed through X.509 certificate infrastructure. It is clear that with LSST's remote computing and data access needs, OAuth token support would greatly simplify access.

The LSST Batch Production service executes campaigns on computing resources to produce the desired LSST data products. A campaign is composed of a set of pipelines, a set of inputs to run the pipelines against, and a method of handling the outputs of the pipelines. A pipeline is a body of code. A campaign is the set of all pipeline executions needed to achieve an LSST objective and pipelines within a campaign can have a dependency chain since outputs from one pipeline might be required as inputs for an following pipeline.

LSST's pipeline orchestration follows a fairly typical pattern. Initially, the best site and computing resources are scheduled or reserved for the pipeline. Pipeline preparation is then run on the remote site and monitored. The pipeline interfaces with LSST's Data Backbone--which serves and manages LSST data products--for required data which then stages the input data and other supporting files into the staging area. It is expected that the Data Backbone will make use of Pegasus \cite{Deelman15} to manage a pipeline's needed data products. The Data Backbone does not include the handling of authorization or authentication of users or services. This functionality is expected to be provided by layers on top of the Data Backbone.

Using HTCondor DAGman \cite{Couvares07}, LSST's pipeline workflow creates a DAG representing pipeline execution. A pipeline would generate multiple condor jobs, comprising the tasks of the pipeline. During execution of the pipeline, a task may require additional data to be staged-in and may require expansion of initial computing resources using ``pilot job'' functionality. As each task ends, any output might be staged from the compute nodes to the Data Backbone or this might not occur until the pipeline concludes. At completion of the pipeline output data is staged out, various custodial tasks concerning the pipeline are executed, a clean up process is executed and the pipeline releases nodes it may have reserved.

SciTokens is being explored for use in LSST's Science Platform.  The LSST Science Platform (LSP) are a set of web applications and services deployed at the LSST Data Access Centers through which the scientific user community will access, visualize, subset and perform analysis of LSST data \cite{LSST1}.  One critical component of the LSP is the enforcement of data access rights.  While LSST data will be made available to the public, there is a period of time when data rights holders will have exclusive access to LSST data for a particular data release \cite{LSST2}.  To that end, the LSP integrates with LSST's identity and access management system to enforce these data access rights.  The LSP has web and database components in which SciTokens could greatly enhance the LSP's ability to authorize user access to data releases.  An example of this would be the LSP aspect that allows user access to data through a RESTful web API.  The LSP would allow users to leverage SciTokens to ensure they are authorized to access the data.

\subsection{OSG}
The Open Science Grid is national, distributed computing partnership for data-intensive research. It allows participating universities and labs to perform distributed, high-throughput computing across heterogeneous resources and at large scale: over 1.3 billion CPU hours were reported in the last year. While the OSG was historically dominated by HEP experiments, it has been rapidly diversifying in the last few years. An important enabler of this diversification is migrating from the legacy X.509-based authentication and authorization layer to more user-friendly mechanisms. Unfortunately, this migration has only been performed for computing jobs: supported storage services still require X.509 authentication. SciTokens is enabling OSG to make progress in storage and data access, an important new capability for the broad set of OSG users.

\subsection{XSEDE}
NSF's Extreme Science and Engineering Discovery Environment (XSEDE) provides a single virtual system that scientists can use to interactively share computing resources, data, and expertise to enhance the productivity of scholars, researchers and engineers. Its integrated, comprehensive suite of advanced digital services is designed to federate with other high-end facilities and with campus-based resources, serving as the foundation for a national e-science infrastructure ecosystem.

Secure distributed access to scientific data is an essential function of the XSEDE infrastructure. Today access control in XSEDE is primarily identity-based, via user accounts, certificates, Duo, and InCommon. The SciToken approach, using OAuth for capability-based (rather than identity-based) access to remote scientific data, can provide new options for XSEDE integration with remote data services (on campus, in the cloud, operated by virtual organizations, etc.). XSEDE has experience using OAuth for web single sign-on to components including CILogon, Globus, MyProxy, and the XSEDE User Portal.

\section{Building Blocks}

SciTokens builds on the software currently in use for LIGO scientific workflows and in many other NSF science projects, which will ease deployment of our work in these environments by updating existing software rather than forcing use of new software. In this section we describe the existing authentication, authorization, and credential management capabilities of HTCondor, CVMFS, and CILogon OAuth that provide the foundational building blocks for SciTokens.

\subsection{HTCondor}
The general architecture of HTCondor \cite{Thain05} is best explained at a high level by understanding three major components. The first major component is known as the "Submit Machine" and is where users typically interface with HTCondor. They submit their jobs and workflows, monitor progress, and see the results of their jobs from this point. The second component is a pool of "Execute Machines" that can vary in size. These are the compute resources which actually execute the users' jobs, and the user does generally not interface with these machines directly. In some environments, users may have access to multiple pools of execute machines that are controlled by different organizations. The third component is the "Matchmaker", which handles the scheduling of jobs and matches them to compute resources according to their requirements, priority of the user, and several other factors. This process runs continuously so that as new jobs enter the queue, they are matched to available resources. Also, if the pool of resources grows (for example, in a cloud computing environment) then again jobs are matched to the newly available resources. When a job is matched, HTCondor spawns a new process on the Submit Machine called the job "Shadow", as well as a new process on the Execute Machine called the job "Starter". These two processes exist for the lifetime of the job and are responsible for moving data, monitoring the job, reporting the status of the job back to the submit machine while the jobs is running, and moving data again when the job has completed.

The HTCondor components communicate with each other over a network using TCP/IP. Sometimes all components of the cluster belong to a single institution and are on a private network behind a firewall and additional security measures are not needed. In other scenarios, the HTCondor components are spread widely across the insecure Internet. HTCondor has a wide array of possible security configurations depending on the requirements of the site and system administrator. In the recommended configuration, communication channels from one HTCondor process to another are authenticated using one of HTCondor's supported mechanisms such as a Shared Secret, Kerberos, or X.509 certificates. Thus, all components of an HTCondor system can trust the other components they interface with. HTCondor also supports encryption of data, both for communication between HTCondor components and also for transferring data as part of running users' jobs. Because the authentication process can be relatively computationally intensive when thousands of nodes are involved, it is desirable not to authenticate every single time a new network connection is made. HTCondor accomplishes this by setting up secure and reusable sessions between components which greatly increases the scalability of the system. For SciTokens, we leverage the secure communication channel between a job's Shadow (on the Submit Machine) and its Starter (on the Execute Machine). It is through this channel that all of a job's executable files, credentials, input, and output data are transmitted. When a large HTCondor system is running, it may be starting dozens of jobs per second, all of which need to establish secure connections between their Shadow and Starter. In this case, the use of sessions does not help. Instead, we rely on a mechanism in which the Matchmaker delegates a trust relationship at the time a job is matched by providing both the Submit and Execute machines with a unique secret key, or ``match password,'' they can use only with each other \cite{Miller10}.

HTCondor currently has a sub-component on the Submit Machine which manages credentials, called the "CredD". The CredD is responsible for securely managing all credentials. The CredD has a plug-in architecture for storing and managing credentials of different types, which we leverage to add support for OAuth tokens as described below. The plug-in architecture includes hooks for refreshing credentials and performing other credential transformations. For example, CERN uses a plug-in to manage Kerberos tickets for jobs and uses also transforms the Kerberos ticket into an AFS token for the running job.

\subsection{CVMFS}
CVMFS (originally, ``CernVM File System'') is a highly-scalable, read-optimized, global distributed filesystem. End-to-end data integrity is achieved through the use of a public key signing and a Merkle-tree-based integrity scheme. CVMFS repositories can consist of millions of files, yet the global system has far more scalable namespaces than traditional cluster filesystem such as Lustre. 

\begin{figure}
\includegraphics[scale=0.35]{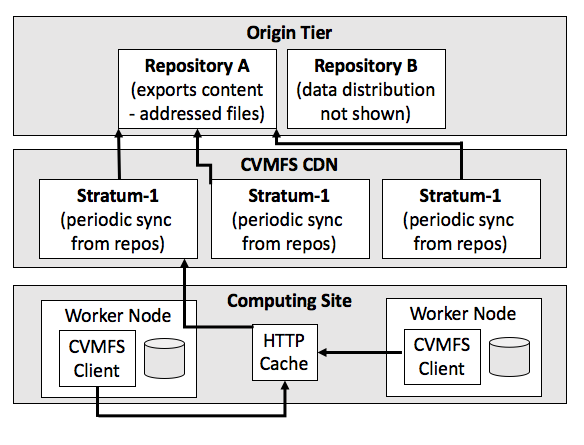}
\caption{CVMFS Architecture}
\label{fig:CVMFSArch}
\end{figure}

This scalability is achieved by performing all writes at one location (the ``repository'') and introducing transaction-based semantics namespace updates. All data and namespace metadata is then content-addressed. The resulting filesystem is extremely amenable to caching techniques; cache hit rates at each layer in the deployed system are often greater than 99\%. Files are distributed using a multi-layer, HTTP-based content distribution network (CDN), site-local HTTP caches, and worker node disk. The traditional CVMFS architecture is outlined in Figure~\ref{fig:CVMFSArch}.

\begin{figure}
\includegraphics[scale=0.35]{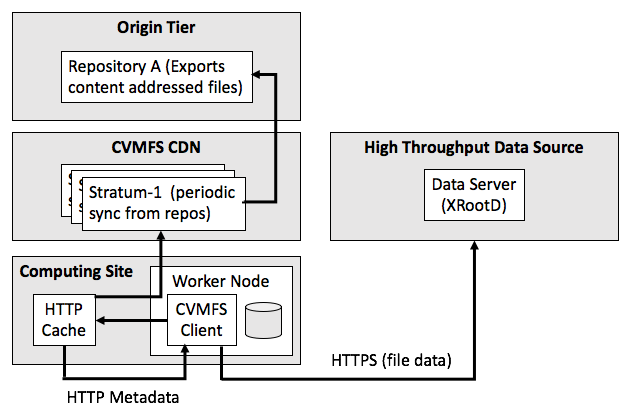}
\caption{Accessing LIGO files in OSG}
\label{fig:CVMFSFed}
\end{figure}

The traditional CVMFS infrastructure works extremely well for distribution of software environments or containers where the data is public and the working set size is one to ten gigabytes (typically, working set size limitations are limited by the HTTP cache and worker node disk size). In 2016, Bockelman and Weitzel implemented a set of changes to the CVMFS client to efficiently handle user authentication / authorization and larger working set sizes \cite{weitzel2017accessing}. This was achieved by allowing CVMFS to access a separate data access infrastructure based upon the XRootD software, utilizing both a high-performance storage system at Nebraska and a series of very large caches run by OSG. This allowed the Nebraska team, in collaboration with the OSG, to configure a CVMFS repository containing the LIGO frame files \cite{weitzel2017data}. Figure~\ref{fig:CVMFSFed} outlines the architecture used to access LIGO files on the OSG.

\begin{figure}
\includegraphics[scale=0.35]{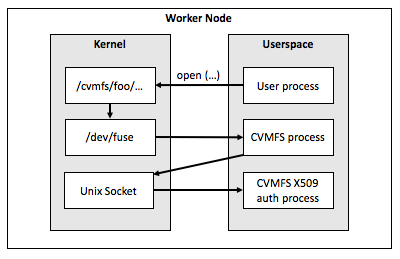}
\caption{Accessing Restricted Data}
\label{fig:CVMFSAuth}
\end{figure}

When a user process accesses a restricted-access repository, the CVMFS process will request an external process to authenticate and authorize the client (the list of authorizations is distributed as an extended attribute in the CVMFS namespace). If the external process can successfully authorize the user process, it will return success and any user credentials back to CVMFS. This authorizes the user to access the local disk cache; if the requested file is not found locally, the user credentials are utilized by the CVMFS client to authenticate the file download from the CVMFS CDN. This is illustrated in Figure~\ref{fig:CVMFSAuth}.

Previously, this plugin API utilized Globus GSI / X.509 certificates for authentication and a list of authorized DNs or VOMS attributes \cite{Alfieri04} for authorization. As described below, the SciTokens project has implemented a new plugin for OAuth tokens.

\subsection{CILogon OAuth}
CILogon \cite{Basney14} provides open source software and an operational service for using federated identities in science projects. CILogon's support for SAML \cite{Cantor05} enables interoperability with the US InCommon federation and other federations worldwide via the eduGAIN interfederation service. CILogon's support for X.509 certificates, compliant with standards from the Interoperable Global Trust Federation (IGTF), enables distributed scientific computing in the Grid Security Infrastructure \cite{Welch03}. Over 4,500 scientists regularly use CILogon for authentication, including over 200 LIGO scientists. CILogon includes support for OAuth \cite{RFC6749} and the OpenID Connect \cite{OIDC} standards, using open source software originally developed for NSF science gateways \cite{Basney11}. This OAuth software contains lightweight Java OAuth client/server libraries, with support for JSON Web Tokens \cite{RFC7519}, which we use for our SciTokens implementation.

\section{Technical Approach}

In this section we present our technical approach for implementing the capability-based tokens for remote data sharing, for LIGO, LSST, and other scientific workflows. Our approach modifies HTCondor to manage the tokens for the workflow, modifies CVMFS to accept the tokens for authorizing remote data access, and implements an OAuth Token Server that issues the capability tokens. In the following subsections, we discuss our technical implementations for each system component.

\subsection{HTCondor}
As the component that actually executes a scientific workflow, HTCondor serves as the linchpin that ties together all the SciTokens components. To best communicate our HTCondor approach, we first present a walk-thru of how HTCondor orchestrates the component interactions upon submission of a job, followed by a discussion of integration points.

\begin{figure}
\includegraphics[scale=0.35]{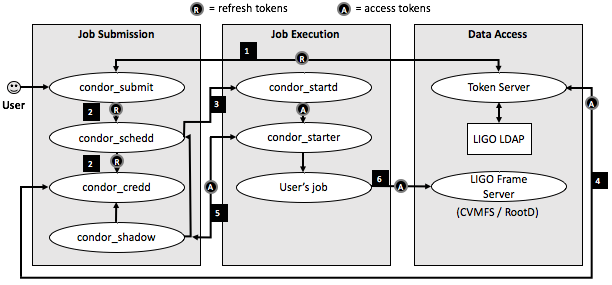}
\caption{The SciTokens System Architecture}
\label{fig:arch}
\end{figure}

As illustrated in Figure~\ref{fig:arch}, the process begins when the researcher submits the computational job using the condor\_submit command (or more likely using Pegasus or similar workflow front-end that then runs condor\_submit). As part of the submission, the researcher specifies required scientific input data and locations for output data storage in the condor\_submit input file. For example, in a LIGO PyCBC \cite{PyCBC16} submission, the researcher will specify a set of data "frames" from the LIGO instrument that are the subject of the analysis. Then condor\_submit authenticates the researcher to the token\_server(s) to obtain the tokens needed for the job's data access; as an optimization, condor\_submit may first check for any locally cached tokens from the researcher's prior job submissions. The token\_server determines if the researcher is authorized for the requested data access, based on the researcher's identity and/or group memberships or other researcher attributes. If the authorization check succeeds, the Token Server issues an OAuth refresh token back to condor\_submit, which stores the refresh token securely in the condor\_credd, and sends the job information to the condor\_schedd. Since condor\_submit gathers all the needed data access tokens, there is no need to store any identity credentials (e.g., passwords, X.509 certificates, etc.) with the job submission, thereby achieving our goal of a capability-based approach.

The next phase of the process begins when the condor\_schedd has scheduled the job on a remote execution site. The condor\_schedd communicates with the condor\_startd to launch the job, establishing a secure communication channel between the condor\_shadow on the submission side and the condor\_starter on the remote execution side. The condor\_starter then requests access tokens from the condor\_shadow for the job's input data. The condor\_shadow forwards the access token requests to the condor\_credd, which uses its stored refresh tokens to obtain fresh access tokens from the token\_server. The condor\_credd returns the access tokens to the condor\_shadow which forwards them securely to the condor\_starter which provides them to the researcher's job. Note that only access tokens are sent to the remote execution environment; the longer-lived refresh tokens remain secured in the submission environment which typically resides at the researcher's home institution. Lastly, the job uses the access tokens to mount CVMFS filesystem(s) to access scientific data. CVMFS verifies each access token to confirm that the token was issued by its trusted token\_server and that the token's scope includes access to the scientific data being requested. If verification succeeds, CVMFS grants the requested data access. If the access token needs to be refreshed, the condor\_starter makes another request back to the condor\_shadow.

Note that SciTokens can leverage additional aspects of HTCondor, such as the fact that the condor\_shadow can be made explicitly aware if the job is staging input data, accessing data online while the job is running, or staging output data. We allow the job submission to state three different sets of access tokens, which will only be instantiated at file stage-in, execution, and file stage-out, respectively. This enables long running jobs, for instance, to fetch a very short-lived write token for output that will only be instantiated once processing has completed. We also adjust the granularity of access token restrictions; for instance, the condor\_shadow may request fresh access tokens for each job instance, allowing the token to be restricted in origin to a specific execution node. Alternatively, for greater scalability, access tokens can be cached at the credd and shared across all condor\_shadow processes serving jobs that need the same data sets. Finally, we are investigating scenarios in which the data service and its accompanying token service is not fixed infrastructure, but instead is dynamically deployed upon execute nodes, perhaps by the workflow itself. In this scenario, the token service could be instantiated with a set of recognized refresh tokens a priori.

\subsection{CVMFS / XRootD}
The CVMFS (client) and XRootD (server) stack was updated to understand the SciTokens authorization model. Fine-grained models afforded by SciTokens, we are able to implement more restrictive access control policies -- allowing reads or writes at individual directory level for groups inside the VO.  This can be done without either CVMFS client or data server needing to know a global identity of the user (as is the case today).

We implemented a new authentication and authorization callout process for CVMFS, based on the experience with our X.509 implementation. This validates the user's authentication token and, as appropriate, authorizes it for use with the CVMFS repository. 

While any HTTPS server implementation can likely be made to work with the OAuth model, we implemented changes to the XRootD server suite to have its HTTPS protocol implementation support bearer authentication and the tokens issued by the SciToken service. This server implementation was selected in order to integrate cleanly with the existing service at Nebraska and provide continuity with the existing X.509-secured LIGO repository. We integrated our token authorization format with XRootD's authorization plugin framework, allowing token-based reading and writing for appropriately enabled XRootD servers.

\subsection{OAuth Tokens}
The distributed, large-scale architecture of CVMFS presents a special challenge for the OAuth access tokens. Tokens are usually validated via the token server's introspection endpoint \cite{RFC7662}; however, the CVMFS client on each worker node must perform access control. This presents a scalability challenge for the LIGO use case. Ideally, the token's verification should be decentralized as opposed to relying on a token service callback. For this reason, we selected JSON Web Tokens \cite{RFC7519} as our token format. These tokens can be verified knowing only the public key of the signing service; CVMFS contains built-in mechanisms for secure public key distribution.

\section{Related Work}

\subsection{ALICE XrootD Tokens}
One of the experiments on the LHC with a comparatively smaller footprint within the US, ALICE never adopted the X.509-credential-based authentication system used throughout the OSG. Instead of using common interfaces (e.g., Globus GridFTP) for data management, ALICE runs private XRootD hosts on top of site storage systems. Having control over an end-to-end VO-specific system allowed ALICE to develop an innovative authorization system. Users would request data access from the central ALICE file catalog; if access was granted, the central service would return an encrypted, base64-encoded XML document describing a list of read / write permissions the bearer is granted. The ALICE-managed XRootD servers would decrypt the token and allow the bearer appropriate authorizations. The data servers would not need the user identity to allow read / write access - effectively, they delegated the management of the ALICE VO's storage allocation to the central ALICE service.

ALICE demonstrated the viability of many of the concepts within SciTokens; we further the approach by:
\begin{itemize}
\item Utilizing the standardized OAuth framework instead of a homegrown format (admittedly, ALICE's work predates these standards by more than 5 years).
\item Investigating token formats that allow decentralized validation: each CVMFS client will need to perform validation, not just the data service.
\item Integrate token generation with federated identity.
\end{itemize}

\subsection{Globus Auth}
Globus Auth \cite{Tuecke16} provides an OAuth-based cloud service for remote scientific data access using GridFTP \cite{Allcock05}, with extensions for delegated access tokens. In contrast to our SciTokens model, Globus Auth has a strong dependency on a centralized, closed-source token service operated by the Globus organization. Tokens in Globus Auth are opaque and require token introspection callbacks to the central service for validation. 

\section{Implementation Status}

All SciTokens code is open source and published at \url{https://github.com/scitokens}.

\subsection{HTCondor}
HTCondor 8.7.7\cite{HTCondor877} includes an initial implementation of OAuth support in the credential management subsystem, using a plug-in model to support different OAuth services (i.e., SciTokens, Box.com, Google Drive, DropBox, etc.).


\subsection{Python Library}
A Python SciTokens reference library \cite{derek_weitzel_2018_1187173} was developed and is being used by both XRootD, CVMFS, and NGINX integrations.  The python library allows the parsing of tokens, as well as testing the token for authorization to access resources.

\subsection{XRootD}
A extension to XRootD was developed that allows XRootD to interpret SciTokens and enforce policies based on those SciTokens \cite{brian_bockelman_2018_1206218}.  XRootD receives connections requests through the HTTPs protocol, which includes the SciToken for authorization.  XRootD uses the extension to verify the SciToken, then provides ACL's that the XRootD subsystem will use for resource access. In addition, a third party transfer extension was developed for XRootD that used SciTokens for authentication.

\subsection{NGINX}
NGINX \cite{reese2008nginx} is a web server that serves 25\% of websites \cite{netcraftNginx}.  An extension was developed \cite{derek_weitzel_2018_1205539} which allows NGINX to verify a Scitoken's validity before allowing access to the resource.  This has been used in combination with NGINX's built-in WebDav \cite{whitehead1998webdav} to create a SciTokens file server.

\subsection{OAuth}
Using our existing Java OAuth client/server libraries, we have developed a SciTokens Authorization Server \cite{scitokens-java} that issues JSON Web Tokens in the SciTokens format, with user and group-based authorization policies for the issued claims (audience, scope, etc.). For example, the server policy can specify that a LIGO user who is a member of the LDGUsers group can obtain a token for reading LIGO frame files and writing to the user's personal output directory.

\section{Conclusions and Next Steps}

The JSON Web Token and OAuth standards provide a solid foundation for distributed, capability-based authorization for scientific workflows. By enhancing existing components (CILogon, CVMFS, HTCondor, XrootD) to support the SciTokens model, we have provided a migration path from X.509 identity-based delegation to OAuth capability-based delegation for existing scientific infrastructures.

Now that we have a software suite that supports the SciTokens model, our next step is to evaluate our end-to-end approach for LIGO and LSST workflows. We will work with the LIGO PyCBC workflow \cite{PyCBC16}, which uses the Pegasus workflow management system \cite{Deelman15} to manage HTCondor job submissions. In addition to demonstrating the use of OAuth tokens for fetching input frame data from CVMFS as part of the LIGO workflows, we will also demonstrate storing PyCBC output data to XrootD and OAuth-capable cloud storage (e.g., Google Cloud Storage), features desired by the PyCBC group.

\begin{acks}
This material is based upon work supported by the National Science Foundation under Grant No.~1738962.
\end{acks}

\bibliographystyle{ACM-Reference-Format}
\bibliography{scitokens}

\end{document}